\documentclass[12pt]{article}
\usepackage{amsfonts}
\usepackage{amsmath, amsthm}
\usepackage{epsfig}
\usepackage{algorithm}

\usepackage{soul}
\usepackage{diagbox}
\usepackage[normalem]{ulem}
\usepackage{amsfonts}
\usepackage{epsfig}
\usepackage{booktabs}
\usepackage{lscape}
\usepackage{multirow}
\usepackage{longtable}
\usepackage{rotating}
\usepackage{array}
\usepackage{multicol}
\usepackage{graphicx}
\usepackage{bigstrut}
\usepackage{setspace}
\usepackage{epstopdf}
\usepackage{mathrsfs}
\usepackage{amsfonts}
\usepackage{epsfig}
\usepackage{booktabs}
\usepackage{threeparttable}
\usepackage{epstopdf}
\usepackage{lscape}
\usepackage{multirow}
\usepackage{threeparttable}
\usepackage{subfigure}

\usepackage{dcolumn}
\newcolumntype{d}[1]{D{.}{.}{#1}}

\usepackage{algorithmic}
\usepackage{amsmath,verbatim,color,amssymb,epsfig}
\usepackage{bm}
\usepackage{amsfonts}
\usepackage{epsfig}
\usepackage{multirow}
\usepackage{graphicx}
\usepackage{array}
\usepackage[table]{xcolor}
\definecolor{Gray}{gray}{0.80}
\usepackage{lscape}
\usepackage{natbib}
\usepackage[normalem]{ulem}
\usepackage[colorlinks=true,urlcolor=blue,citecolor=black,linkcolor=blue,bookmarks=true]{hyperref}
\usepackage{caption}
\begin{document}
\def\eqx"#1"{{\label{#1}}}
\def\eqn"#1"{{\ref{#1}}}

\makeatletter 
\@addtoreset{equation}{section}
\makeatother  

\def\yincomment#1{\vskip 2mm\boxit{\vskip 2mm{\color{red}\bf#1} {\color{red}\bf --Yin\vskip 2mm}}\vskip 2mm}
\def\squarebox#1{\hbox to #1{\hfill\vbox to #1{\vfill}}}
\def\boxit#1{\vbox{\hrule\hbox{\vrule\kern6pt
          \vbox{\kern6pt#1\kern6pt}\kern6pt\vrule}\hrule}}

\def\zhangcomment#1{\vskip 2mm\boxit{\vskip 2mm{\color{blue}\bf#1} {\color{blue}\bf --Zhang\vskip 2mm}}\vskip 2mm}
\def\squarebox#1{\hbox to #1{\hfill\vbox to #1{\vfill}}}
\def\boxit#1{\vbox{\hrule\hbox{\vrule\kern6pt
          \vbox{\kern6pt#1\kern6pt}\kern6pt\vrule}\hrule}}

\newcommand{\blue}[1]{\textcolor{blue}{{#1}}}
\newcommand{\red}[1]{\textcolor{red}{{#1}}}
\def\theequation{\thesection.\arabic{equation}}
\newcommand{\ds}{\displaystyle}

\newcommand{\bJ}{\mathbf{J}}
\newcommand{\bF}{\mathbf{F}}
\newcommand{\bM}{\mathbf{M}}
\newcommand{\bR}{\mathbf{R}}
\newcommand{\bZ}{\mathbf{Z}}
\newcommand{\bX}{\mathbf{X}}
\newcommand{\bx}{\mathbf{x}}
\newcommand{\bQ}{\mathbf{Q}}
\newcommand{\bH}{\mathbf{H}}
\newcommand{\bh}{\mathbf{h}}
\newcommand{\bz}{\mathbf{z}}
\newcommand{\ba}{\mathbf{a}}
\newcommand{\be}{\mathbf{e}}
\newcommand{\bG}{\mathbf{G}}
\newcommand{\bB}{\mathbf{B}}
\newcommand{\bb}{\mathbf{b}}
\newcommand{\bA}{\mathbf{A}}
\newcommand{\bC}{\mathbf{C}}
\newcommand{\bI}{\mathbf{I}}
\newcommand{\bD}{\mathbf{D}}
\newcommand{\bU}{\mathbf{U}}
\newcommand{\bc}{\mathbf{c}}
\newcommand{\bd}{\mathbf{d}}
\newcommand{\bs}{\mathbf{s}}
\newcommand{\bS}{\mathbf{S}}
\newcommand{\bV}{\mathbf{V}}
\newcommand{\bv}{\mathbf{v}}
\newcommand{\bW}{\mathbf{W}}
\newcommand{\bw}{\mathbf{w}}
\newcommand{\bg}{\mathbf{g}}
\newcommand{\bu}{\mathbf{u}}
\newcommand{\by}{\mathbf{y}}
\def\bb{{\bf b}}
\newcommand{\cov}{\mbox{cov}}
\newcommand{\var}{\mbox{var}}
\newcommand{\diag}{\mbox{diag}}

\newcommand{\bepsilon}{\boldsymbol{\epsilon}}
\newcommand{\bvarepsilon}{\boldsymbol{\varepsilon}}
\newcommand{\bcU}{\boldsymbol{\cal U}}
\newcommand{\bbeta}{\boldsymbol{\beta}}
\newcommand{\bdelta}{\boldsymbol{\delta}}
\newcommand{\bDelta}{\boldsymbol{\Delta}}
\newcommand{\boldeta}{\boldsymbol{\eta}}
\newcommand{\bxi}{\boldsymbol{\xi}}
\newcommand{\bGamma}{\boldsymbol{\Gamma}}
\newcommand{\bSigma}{\boldsymbol{\Sigma}}
\newcommand{\balpha}{\boldsymbol{\alpha}}
\newcommand{\bOmega}{\boldsymbol{\Omega}}
\newcommand{\btheta}{\boldsymbol{\theta}}
\newcommand{\bmu}{\boldsymbol{\mu}}
\newcommand{\bnu}{\boldsymbol{\nu}}
\newcommand{\bgamma}{\boldsymbol{\gamma}}
\newcommand{\tr}{\mbox{tr}}
\newcommand{\mse}{\mbox{MSE}}

\newtheorem{thm}{Theorem}
\newtheorem{lem}{Lemma}
\newtheorem{rem}{Remark}
\newtheorem{cor}{Corollary}
\newcolumntype{L}[1]{>{\raggedright\let\newline\\\arraybackslash\hspace{0pt}}m{#1}}
\newcolumntype{C}[1]{>{\centering\let\newline\\\arraybackslash\hspace{0pt}}m{#1}}
\newcolumntype{R}[1]{>{\raggedleft\let\newline\\\arraybackslash\hspace{0pt}}m{#1}}

\newcommand{\tabincell}[2]{\begin{tabular}{@{}#1@{}}#2\end{tabular}}

\newcommand{\dT}{\top}

\newcommand{\algorithmicobs}{\textbf{Observations:}}
\newcommand{\algorithmicprior}{\textbf{Prior:}}
\newcommand{\PRIOR}{\item[\algorithmicprior]}
\newcommand{\OBS}{\item[\algorithmicobs]}

\newcommand{\algorithmicoutput}{\textbf{Output:}}
\newcommand{\OUTPUT}{\item[\algorithmicoutput]}

\baselineskip=24pt
\begin{center}
{\Large \bf
PCA Rerandomization
}
\end{center}

\vspace{2mm}
\begin{center}
{\bf Hengtao Zhang$^{1}$, Guosheng Yin$^{1}$ and Donald B. Rubin$^{2}$}
\end{center}

\begin{center}

$^{1}$Department of Statistics and Actuarial Science\\
The University of Hong Kong\\
Pokfulam Road, Hong Kong\\

$^{2}$Department of Statistics\\
Harvard University\\
Cambridge, MA, USA\\
$^{2}$Yau Mathematical Sciences Center \\
Tsinghua University\\
Beijing, China\\
$^{2}$Fox Business School \\
Temple University\\
Philadelphia, PA, USA\\

\vspace{2mm}


\end{center}
\noindent{Abstract.}
Mahalanobis distance between treatment group and control group covariate means is often adopted as a balance criterion
when implementing a rerandomization strategy. However,
this criterion may not work well for high-dimensional cases because it balances all orthogonalized covariates equally.
Here, we propose leveraging principal component analysis (PCA) to identify proper subspaces in which Mahalanobis distance should be calculated. Not only can PCA effectively reduce the dimensionality for high-dimensional cases while capturing most of the information in the covariates, but
it also provides computational simplicity by focusing on the top orthogonal components. We show that our
PCA rerandomization scheme has desirable theoretical properties on balancing covariates and thereby on improving the estimation of average treatment effects. We also show that this conclusion is supported by numerical studies using both simulated and real examples.

\vspace{0.5cm}

\vspace{1cm}
\section{Introduction}

Randomized experiments have long been regarded as the gold standard to measure the effect of an intervention, because randomization can reduce the potential bias of estimates by balancing the covariate distributions between treatment groups on average. However, when pure (complete) randomization is implemented in practice, it often yields unbalanced allocations, so that the groups should be rerandomized before the experiment is actually conducted. Although rerandomization has been discussed earlier  \citep{fisher1926arrangement,cox2009randomization,worrall2010evidence}, its formal theoretical framework was not well established until the publication of \cite{morgan2012rerandomization}. Using Mahalanobis distance as the balance criterion for treatment-control experiments, rerandomization was shown to improve the covariate balance and the precision of estimated treatment effects. Following the work of \cite{morgan2012rerandomization}, effort has been made to extend or modify such rerandomization schemes. For example, \cite{morgan2015rerandomization} proposed a rerandomization strategy for covariates with different tiers of anticipated importance with respect to the outcome variable. The extension of rerandomization to a $2^K$ factorial design was developed by \cite{branson2016improving} based on a real example with educational data. \cite{zhou2018sequential} considered rerandomization for experiments with sequentially enrolled units. \cite{li2018asymptotic,li2020rerandomizationfactorial} investigated the asymptotic properties of the standard treatment effect estimator for the treatment-control settings and $2^K$ factorial designs respectively. \cite{li2020rerandomization} further established asymptotic properties for the combination of rerandmization and regression adjustment.

All the aforementioned works use Mahalanobis distance as the balance measure owing to its several appealing characteristics. First, it is invariant to any affine transformation of the original covariates. Second, \cite{morgan2012rerandomization} showed that, not only can Mahalanobis distance guarantee the unbiasedness of the treatment effect estimator as well as the balance of covariate means between equal-sized treatment and control groups, but it also reduces an equal percent of sampling variance for each covariate and principal component. Apart from rerandomization, Mahalanobis distance is also widely applied in matching methods for observational studies \citep{rubin1973matching,rubin1973use, rubin1979using,rubin1980bias,rosenbaum1985constructing,stuart2010matching}.

Despite the advantages discussed above, the full-rank Mahalanobis distance may not work well for rerandomization with the high-dimensional data \citep{branson211ridge}. Because it is difficult to equally balance a large number of principal components with different magnitudes of sampling variances.
In related work, \cite{morgan2015rerandomization}
proposed balancing the covariates hierarchically using prespecified tiers of importance of covariates
related to the outcome. \cite{branson211ridge} pointed out that it might be
difficult to specify the relative importance for a large number of covariates {\em a priori}.
They proposed including a ridge term in Mahalanobis distance, which they showed puts more emphasis on the top principal components.
However, the ridge rerandomization of
\cite{branson211ridge} relies on complicated Monte Carlo integration and constraint optimization to determine the value of the ridging parameter. 
Rather than using Mahalanobis distance, \cite{johansson2020rerandomization} proposed a different rank-based balance measure for rerandomization, but their heuristic metric is designed for longitudinal data where the pre-experimental outcomes are available to
estimate the relative importance of each covariate. Moreover, the theoretical properties are have not yet been developed under their proposed balance metric.

Here, we propose using only the top principal components from a principal component analysis (PCA) to calculate Mahalanobis distance in the associated subspace and then perform rerandomization. Our PCA rerandomization can be viewed as using a lower-dimensional alternative to the full-rank Mahalanobis distance. We show that because PCA rerandomization only reduces the variance of
the selected top components, it thereby
imposes more shrinkage on them relative to full-rank rerandomization given the same acceptance probability. 
Moreover, the lower dimensional orthogonality of top principal components simplifies the covariance matrix into a diagonal matrix and thus improves the computational efficiency when calculating Mahalanobis distance.

We
establish theory for PCA rerandomization, including the sampling distribution of the modified balance
criterion and the variance reduction properties for the standard treatment effect estimator and covariate mean differences
compared with the complete randomization. Practically, despite using PCA, our method is as easy to implement as the
original rerandomization
without cumbersome parameter specification or increased computation involved in the ridge rerandomization.

In Section \ref{sec:rer_review}, we review the rerandomization framework based on Mahalanobis distance \citep{morgan2012rerandomization}. We present the details of PCA rerandomization and its theoretical properties in Section \ref{sec:pcarer}. Section \ref{sec:numstudy} reports the  results of numerical experiments, which show the potential desirable performance of our proposed method compared with other randomization schemes. Section \ref{sec:discuss} concludes with a discussion.

\section{Rerandomization with Mahalanobis Distance}\label{sec:rer_review}

Let $\bX=(\bx_1,\dots,\bx_n)^{\top}\in\mathbb{R}^{n\times d}$ be the fixed covariate matrix representing $n$ trial participants with $\bx_i\in\mathbb{R}^{d}$. For simplicity, assume that all $\bx_i$'s are standardized to have zero mean and unit variance: $\bX^\top\bm{1}_n=\mathbf{0}$, where $\bm{1}_n\in\mathbb{R}^n$ is the $n$-component vector with all components equal to 1. We focus here on treatment-versus-control experiments, and define $\bW=(W_1,\dots, W_n)^{\top}\in\{0,1\}^{n}$ to be the random indicator vector of allocations, where $W_i=1$ means
the $i$th unit is assigned to the treatment group, whereas $W_i=0$ means the $i$th unit is assigned to the control group. In the theoretical developments, for simplicity, we further impose the constraint on $\bW$ such that treatment and control groups are the same size,
\begin{align}\label{def: nt=nc}
\sum_{i=1}^{n}W_i=\sum_{i=1}^{n}(1-W_i)=n/2.
\end{align}
Given $\bW$ and $\bX$, let
\begin{align}\label{def: xt&xc}
\bar{\bx}_{T} = \frac{2}{n}\bX^{\top}\bW \quad \mbox{and} \quad \bar{\bx}_{C} = \frac{2}{n}\bX^{\top}({\bf 1}_n-\bW),
\end{align}
which denote the mean vector of the covariates of the treatment and control groups respectively. According to the potential outcomes framework \citep{neyman1923edited,rubin1974estimating,imbens2015causal}, each unit is associated with two potential outcomes $y_i(1)$ or $y_i(0)$ corresponding to the situation where the unit is assigned to the treatment group or the control group. Only one outcome can be observed after the allocation, which can be written as $y_{i}=W_iy_i(1)+(1-W_i)y_i(0)$.  The standard goal of causal inference is to estimate the sample average treatment effect (SATE),
$$\tau = \frac1n\sum_{i=1}^n\{y_i(1)-y_i(0)\}.$$
The standard simple estimator of SATE is the mean difference of the observed outcomes between the two observed groups,
$$\hat{\tau} = \bar{y}_T-\bar{y}_C = \frac2n\big\{\bW^{\top}\by - (\mathbf{1}_n-\bW)^{\top}\by\big\},$$
where $\by=(y_{1},\dots,y_{n})^{\top}$.
In randomized experiments, more balanced covariates across treatment and control groups generally lead to a more precise estimator $\hat{\tau}$. Traditional pure (or complete) randomization only balances covariates between the treatment and control groups on average, and thus a particular realized allocation can be unbalanced, thereby adversely affecting statistical inference.

\cite{morgan2012rerandomization} formally established the rerandomization framework using
Mahalanobis distance, $M$, as the balance measure, where
\begin{align}\label{def:mdist}
M = (\bar{\bx}_T-\bar{\bx}_C)^{\top}\bSigma^{-1}(\bar{\bx}_T-\bar{\bx}_C),
\end{align}
and $\bSigma=\cov(\bar{\bx}_T-\bar{\bx}_C|\bX)=4\cov(\bX)/n$ is the covariance matrix of $\bar{\bx}_T-\bar{\bx}_C$ with respect to all $\bW$ satisfying condition (\ref{def: nt=nc}), and $\cov(\bX)=\bX^{\top}\bX/(n-1)$ is the sample covariance matrix of $\bX$. When $\bSigma$ is singular, the pseudo-inverse is applied to calculate the distance. Rather than performing one single randomization, basic rerandomization continues generating feasible allocation vectors $\bW$ until the corresponding $M$ is smaller than a predefined threshold $a$ ($a>0$). The first $\bW$ with $M\leq a$ is chosen to determine the actual allocation in the experiment. Under the mild condition that $\bar{\bx}_T-\bar{\bx}_C | \bX \sim N(\bm{0}, \bSigma)$, the distribution of Mahalanobis distance $M | \bX$ follows a $\chi^2_d$ distribution, so the threshold $a$ can be determined by controlling the acceptance probability $p_a$ with $P(M\leq a|\bX)=p_a$.

Rerandomization using Mahalanobis distance has several attractive properties. First, rerandomization ensures the unbiased estimation of $\tau$, that is, $\mathbb{E}(\hat{\tau}|\bX, M\leq a)=\tau$, because it balances the mean difference of any observed or even unobserved covariate $x$: $\mathbb{E}(\bar{x}_T-\bar{x}_C|\bX, M\leq a)=0$. Second, it reduces the variance of each covariate by the same proportion, $100(1-v_a)\%$ relative to complete randomization,
\begin{align}\label{def:evpr}
\cov(\bar{\bx}_T-\bar{\bx}_C|\bX,M\leq a) = v_a \cov(\bar{\bx}_T-\bar{\bx}_C|\bX),
\end{align}
where $v_a = P(\chi^2_{d+2}\leq a)/P(\chi^2_{d}\leq a) \in (0,1)$. Similarly, if there exists a linear relationship between the outcome $y$ and covariate $\bx$, the variance reduction for the estimated treatment effect, $\var(\hat{\tau}|\bx, M\leq a) = \left\{1-(1-v_a)R^2\right\}\var(\hat{\tau}|\bx)$, where $R^2$ is the multiple squared correlation between $y$ and $\bx$.

\cite{branson211ridge} developed ridge rerandomization by including a ridge term when calculating Mahalanobis distance,
\begin{align*}
M_{\lambda} = (\bar{\bx}_T-\bar{\bx}_C)^\top(\bSigma+\lambda\bI_d)^{-1}(\bar{\bx}_T-\bar{\bx}_C),
\end{align*}
where $\bI_d$ denotes the $d$-dimensional identity matrix; they showed that such ridging automatically assigns more weight to the top components and can provide better balance in high-dimensional/high-collinearity cases. 
We propose using the top principal components rather than original covariates (i.e., all principal components) to perform the rerandomization. Our lower-dimensional method inherits most of the theoretical properties of rerandomization and can be more convenient to implement without possibly cumbersome specifications of tuning parameters as in the ridge rerandomization.

\section{PCA Rerandomization}\label{sec:pcarer}
\subsection{Rerandomization using Principal Components}
Define $\bX=\bU\bD\bV^{\top}$ to be the singular value decomposition of $\bX$, where $\bU\in\mathbb{R}^{n\times p}$ and $\bV\in\mathbb{R}^{d\times p}$ correspond to the matrices of the left and right singular vectors, with $\bU^{\top}\bU=\bV^{\top}\bV=\bI_p$ and $p=\min\{n,d\}$;
 $\bD=\diag\{\sigma_1,\dots,\sigma_p\}$ is a diagonal matrix composed of non-negative singular (i.e., eigen) values $\sigma_1\geq\dots\geq\sigma_p>0$. Here, we focus on the common case with $p=d$ 
 so that $\bZ=(z_{ij})=\bU\bD$ are the principal components of $\bX$. Let $\bZ_k=\bU_k\bD_k=\big(\bz^{(k)}_{1},\dots,\bz^{(k)}_{n}\big)^{\top}\in\mathbb{R}^{n\times k}$ denote the matrix of
the top $k$ ($k\leq d$) principal components of $\bX$, where $\bz^{(k)}_i$ is the first $k$ elements of the $i$th row of $\bZ$, and $\bU_k\in\mathbb{R}^{n\times k}$ is the first $k$ columns of $\bU$ and $\bD_k\in\mathbb{R}^{k\times k}$ is the top $k$-dimensional sub-matrix of $\bD$. Similarly, let $\tilde{\bZ}_{d-k}=\tilde{\bU}_{d-k}\tilde{\bD}_{d-k}\in\mathbb{R}^{n\times (d-k)}$ denote the last $d-k$ principal components, where $\tilde{\bU}_{d-k}\in\mathbb{R}^{n\times (d-k)}$ is the last $d-k$ columns of $\bU$ and $\tilde{\bD}_k\in\mathbb{R}^{(d-k)\times (d-k)}$ is the last $(d-k)$-dimensional sub-matrix of $\bD$. We calculate Mahalanobis distance based on the top $k$ principal components,
\begin{align}\label{def:mdist_pca}
M_{k} = \big(\bar{\bz}^{(k)}_T-\bar{\bz}_C^{(k)}\big)^{\top}\bSigma_{z}^{-1}\big(\bar{\bz}^{(k)}_T-\bar{\bz}_C^{(k)}\big),
\end{align}
where $\bar{\bz}^{(k)}_T$ and $\bar{\bz}^{(k)}_C$ are defined similarly following (\ref{def: xt&xc}),
$$\bar{\bz}^{(k)}_{T} = \frac{2}{n}\bZ_k^{\top}\bW \quad \mbox{and} \quad \bar{\bz}^{(k)}_{C} = \frac{2}{n}\bZ_k^{\top}({\bf 1}_n-\bW),$$
and $\bSigma_{z}=C_n\bZ_k^{\top}\bZ_k=C_n\bD_k^2$ with
$C_n = 4/(n^2-n)$. For selecting a treatment assignment,
we proceed by generating $\bW$ until the criterion $M_{k}\leq a_k$ is reached, which is referred to as PCA rerandomization, or more precisely PCA-$k$ rerandomization.

For a particular randomized allocation, let $\bar{z}_{T,j}$ and $\bar{z}_{C,j}$ be the mean values of the $j$th principal component for the treatment and control groups respectively. Let $s^2_j$ denote the sample variance of the $j$th component. It can be shown that $s^2_j=\sum_{i=1}^nz^2_{ij}/(n-1)=\sigma_j^2/(n-1)$, because $\bZ$ is also centered, $\bm{1}_n^\top\bZ=\bm{1}_n^\top\bX\bV=\bm{0}$. Thus, we can rewrite (\ref{def:mdist_pca}) as
\begin{align*}
M_{k} = \frac{n}{4} \sum_{j=1}^k \left(\frac{\bar{z}_{T,j}-\bar{z}_{C,j}}{s_j}\right)^2,
\end{align*}
which is the sum of standardized mean difference of the top $k$ principal components. Note that $\bZ$ can be obtained from the affine transformation on covariates $\bZ = \bX\bV$. Following the affinely invariant property of Mahalanobis distance,  it can be shown that
\begin{align*}
M =  \frac{n}{4} \sum_{j=1}^d \left(\frac{\bar{z}_{T,j}-\bar{z}_{C,j}}{s_j}\right)^2~~\mbox{and}~~M_\lambda =  \frac{n}{4} \sum_{j=1}^d \frac{\sigma^2_j}{\sigma^2_j+\lambda/C_n}\left(\frac{\bar{z}_{T,j}-\bar{z}_{C,j}}{s_j}\right)^2,
\end{align*}
where the former is the original Mahalanobis distance and the latter corresponds to the ridge rerandomization.
Therefore, PCA rerandomization is a truncated version of the original rerandomization, i.e., $M_k\leq M$ because of $k\leq d$ where the equality holds only if $k=d$. Furthermore, $M_\lambda$ is a weighted version of $M$, with weight $\sigma_j^2/(\sigma_j^2+\lambda/C_n)$ attached to all components, i.e., there is no dimension reduction. Therefore, ridge rerandomization can be viewed as a smooth counterpart to PCA rerandomization,  which uses a binary weight, $$1_{j\leq k}(j) = \begin{cases}
\sigma_j^2/(\sigma_j^2+0)=1, & \mbox{if} ~ j\leq k,\\
\sigma_j^2/(\sigma_j^2+\infty)=0, & \mbox{if} ~ j> k.
\end{cases}$$

Although there are two tuning parameters for PCA rerandomization, they are easy to specify. As in the routine PCA, the number of top components $k$ can be determined using the percent of cumulative variation that is explained by the selected principal components,
\begin{align}\label{def:pcak}
k=\min_j\left\{j\middle\vert\frac{\sum_{i=1}^j\sigma^2_i}{\sum_{i=1}^d\sigma^2_i}\geq \gamma_k\right\},
\end{align}
where $\gamma_k\in(0,1)$ is a prespecified constant.
Given $k$, we determine the threshold $a_k$ via the acceptance probability $p_{a_k}$ analogous to the full-rank rerandomization, $P(M_k\leq a_k|\bX)=p_{a_k}$, where the distribution of $M_k|\bX$ is $\chi^2_k$.

\subsection{Theoretical Properties of PCA Rerandomization}\label{subsec:property_pcarer}
Given $k$ and $a_k$, several statistical properties can be derived for the PCA rerandomization, and we defer the corresponding technical details to the Appendix. PCA rerandomization balances the covariates between the treatment and control groups on average, and additionally leads to the unbiased estimation of $\tau$. 

\begin{thm}\label{thm:unbias}
Given a constant $a_k>0$,
$$\mathbb{E}\left(\bar{\bx}_T-\bar{\bx}_C|\bX, M_{k}\leq a_k\right)=0\quad\mbox{and} \quad
\mathbb{E}\left(\hat{\tau}|\bX,M_{k}\leq a_k\right) = \tau.
$$
\end{thm}
According to the definition of $M_{\rm k}$, it has the same value
for both allocations $\bW$ and $\mathbf{1}_n-\bW$ for any given threshold $a_k>0$. Furthermore, we assume that equation (\ref{def: nt=nc}) holds in PCA rerandomization. Therefore,
the unbiasedness for $\bar{\bx}_T-\bar{\bx}_C$ and $\hat{\tau}$ follows
Theorem 2.1 and Corollary 2.2 in \cite{morgan2012rerandomization}.

The characteristic of covariate balance in Theorem \ref{thm:unbias} can be further extended to the unobserved covariates, as implied by Corollary 2.2 of \cite{morgan2012rerandomization}. In addition to removing the conditional bias, PCA rerandomization tends to make the difference of covariate means and $\hat{\tau}$ more concentrated. Before showing such results, we first provide the distribution of $M_k$ under the condition $(\bar{\bx}_T-\bar{\bx}_C)|\bX\sim N(\bm{0},\bSigma)$.

\begin{thm}\label{thm:dist_Mpca}
Let $k$ denote the number of selected top principal components. If $(\bar{\bx}_{T}-\bar{\bx}_{C})|\bX\sim N(\mathbf{0},\bSigma)$, we have that
$$M_{k}|\bX\sim \chi^2_k.$$
\end{thm}

Under PCA rerandomization, $M_k$ thus follows a chi-squared distribution with the degree of freedom being the same as the number of selected top principal components. The original rerandomization yields $M|\bX\sim \chi^2_d$,  because it uses all $d$ components. One major application of Theorem \ref{thm:dist_Mpca} is to specify the threshold $a_k$ through the probability $p_{a_k}$ such that $p_{a_k} = P(\chi^2_k\leq a_k)$. Smaller degree of freedom yields a smaller threshold given the same acceptance probability $p_a=p_{a_k}$, so we have $a_k<a$ for PCA rerandomization relative to the threshold in pure rerandomization. Moreover, it is easier for our method to determine $a_k$ than ridge rerandomizaiton, which has to specify the threshold from a mixture distribution. Given the distribution of $M_k$ and the threshold $a_k$, we can quantify how much variance is reduced for the covariance matrix of $\bar{\bx}_T-\bar{\bx}_C$ through the following theorem.

\begin{thm}\label{thm:cov_reduce}
Given the top $k$ principal components and the threshold $a_k>0$, if $(\bar{\bx}_T-\bar{\bx}_C)|\bX\sim N(\mathbf{0},\bSigma)$,
$${\rm cov}(\bar{\bx}_T-\bar{\bx}_C|\bX, M_{k}\leq a_k)=C_n\bV\begin{pmatrix}
v_{a_k}\bD^2_k & \mathbf{0}\\
\mathbf{0} & \tilde{\bD}^2_{d-k}
\end{pmatrix}\bV^{\top},$$
where $C_n=4/(n^2-n)$ and $v_{a_k} = P(\chi^2_{k+2}\leq a_k)/P(\chi^2_{k}\leq a_k)$.
\end{thm}

Theorem \ref{thm:cov_reduce} states the obvious fact that PCA rerandomization only reduces the variation of top $k$ selected principal components, for which the percent reduction in variance (PRV) is $100(1-v_{a_k})\%$. This strategy automatically identifies and balances the \textit{most variable subspace} of original covariates. Furthermore, we find that $v_{a_k}$ in PCA rerandomization is at most its rerandomization counterpart $v_a$ given the same acceptance probability.

From \cite{morgan2012rerandomization}, the covariance reduction for rerandomization is
\begin{align}\label{eqn:cov_reduce_rer}
\cov(\bar{\bx}_T-\bar{\bx}_C|\bX, M\leq a)=C_n\bV\begin{pmatrix}
v_a\bD^2_k & \mathbf{0}\\
\mathbf{0} & v_a\tilde{\bD}^2_{d-k}
\end{pmatrix}\bV^{\top}.
\end{align}
Therefore, smaller $v_{a_k}$ means that we reduce more variance along the top principal axes than pure rerandomization. 
In contrast to ridge rerandomization, our scheme only reduces the same percent of variance for the selected $k$ components, whereas ridge rerandomization reduces for all components but with different percents, with top components receiving more shrinkage, i.e.,
$$\cov(\bar{\bx}_T-\bar{\bx}_C|\bX, M_\lambda\leq a_\lambda)=C_n\bV\diag\left(\xi_{\lambda,1}\sigma_1^2,\dots,\xi_{\lambda,d}\sigma_d^2\right)\bV^{\top},$$
where $0<\xi_{\lambda,1}\leq \dots \leq \xi_{\lambda,d}$ are constants defined in Theorem 4.2 of \cite{branson211ridge} for a given $\lambda$.
The aforementioned discussion focuses on the PRV on principal components, whereas
the following corollary further provides the PRV for the $j$th entry of the covariate mean difference $\bar{x}_{T,j}-\bar{x}_{C,j}$ under PCA rerandomization.

\begin{cor}\label{cor: cov_reduce}
Given the conditions in Theorem \ref{thm:cov_reduce} and 
the $(j,j)$th element of $\bSigma$, $\bSigma_{jj}>0$, for $j\in[d]=\{1,\dots,d\}$, the PRV of $\bar{x}_{T,j}-\bar{x}_{C,j}$ is $100(1-v_{a_k,j})\%$ for $j\in[k]$, where
$$v_{a_k,j} = \frac{\big(C_n\bV{\rm diag}\big\{v_{a_k}\bD^2_k,\tilde{\bD}^2_{d-k}\big\}\bV^{\top}\big)_{jj}}{\bSigma_{jj}}\in(0,1).$$
\end{cor}

To assess how PCA rerandomization can improve the
estimation of $\tau$, we follow \cite{morgan2012rerandomization} to assume that the
treatment effect is additive:
\begin{align}\label{def:taumodel}
y_i(W_i) = \beta_0+\bx^{\top}_i\bbeta+\tau W_i+\varepsilon_i,
\end{align}
where $\beta_0+\bx^{\top}_i\bbeta$ is the projection of the outcome $y_i$ onto the subspace spanned by $(\mathbf{1},\bX)$, and $\varepsilon_i$ is the residual of $y_i$ orthogonal to the linear subspace spanned by $\bX$. We further assume that $\hat{\tau}$, the estimator of $\tau$, follows a normal distribution given $\bX$, as in \cite{morgan2012rerandomization}.

\begin{thm}\label{thm: tau_reduce}
Given the conditions in Theorem \ref{thm:cov_reduce}, if the data arise from (\ref{def:taumodel}) and $\hat{\tau}$ is normally distributed given $\bX$, then for any $\bbeta\in\mathbb{R}^{d}$,
$${\rm var}(\hat{\tau}|\bX) - {\rm var}(\hat{\tau}|\bX, M_{k} \leq a_k) = C_n\bbeta^{\top}\bV\begin{pmatrix}
(1-v_{a_k})\bD^2_k & \mathbf{0}\\
\mathbf{0} & \mathbf{0}
\end{pmatrix}\bV^{\top}\bbeta\geq 0.$$
Furthermore, let $\tilde{\bbeta}=\bV^{\top}\bbeta$ where the $j$th component of $\tilde{\bbeta}$ is denoted by $\tilde{\beta}_j$, and $\mathcal{C}=\{\bbeta\in\mathbb{R}^d|\exists j\in[k]~\mbox{s.t.}~\tilde{\beta}_j\neq 0\}$. If $\bbeta\in \mathcal{C}$,
$${\rm var}(\hat{\tau}|\bX) - {\rm var}(\hat{\tau}|\bX, M_{k} \leq a_k)>0. $$
\end{thm}

This theorem shows that PCA rerandomization never harms the precision of
the treatment effect estimator $\hat{\tau}$ relative to pure rerandomization, although it requires an additional constraint $\bbeta\in\mathcal{C}$ to guarantee a strict sampling variance reduction. Note that each column of $\bV$ corresponds to a direction of the principal component.  The constraint intuitively means that the coefficients $\bbeta$ should not fall into the subspace spanned by the last $d-k$ components. As shown in the proof in the Appendix, the sampling variance reduction of $\hat{\tau}$ is related to that of covariates. In particular, we have for all rerandomization schemes,
\begin{align}\label{eqn:vartau_varcov}
{\rm var}(\hat{\tau}|\bX) - {\rm var}(\hat{\tau}|\bX, M\leq a) = \bbeta^\top\big\{\cov(\bar{\bx}_T-\bar{\bx}_C|\bX) - \cov(\bar{\bx}_T-\bar{\bx}_C|\bX, M\leq a)\big\}\bbeta.
\end{align}
Therefore, the variance reduction of $\hat{\tau}$ is governed by all principal components for pure rerandomization and ridge rerandomization due to the fact that  $\cov(\bar{\bx}_T-\bar{\bx}_C|\bX)>\cov(\bar{\bx}_T-\bar{\bx}_C|\bX, M\leq a)$ under both schemes. Specifically, each principal component has an equal contribution to the reduction under pure rerandomization, whereas top principal components are associated with larger weights in $M_\lambda$ and thus result in more variance reduction than the tail components under ridge rerandomization. The reductions of all principal components are finally aggregated through the inner product with the coefficients $\bbeta$. On the other hand, using only the top principal components for PCA rerandomization leads to a constraint to ensure the strict reduction in (\ref{eqn:vartau_varcov}). 

\section{Numerical Studies}\label{sec:numstudy}

\subsection{Simulation Settings}\label{sec:simustudy}
Our Monte Carlo study can be compactly described as a $4\times 4 \times 3 \times 3\times 2 \times 2 \times 2$ factorial design.
The factors can be divided into two categories at the design stage of causal inference following the setups in \cite{rubin1979using} and
\cite{gutman2013robust,gutman2015estimation,gutman2017estimation}. The first type of factor includes the
characteristics of covariate distributions, sample size and rerandomization schemes, which are explicitly known to the investigator or can be estimated without
using the outcome data, and thus are essentially known to the investigator at the design stage. The second type
of factor contains the information about the response surfaces including the variance of residuals,
which are empirically inestimable without outcome data, and therefore generally unknown at the design stage.

We generate $n\in\{100,200,500,1000\}$ samples from a $d$-dimensional multivariate normal distribution, $\bx\sim N\left(\mathbf{0},(1-\rho)\bI_d+\rho\bm{1}_d\bm{1}_d^\top\right),$
where the correlation coefficient $\rho\in\{0.1,0.5,0.9\}$
and the dimension $d\in\{10,50,90,180\}$. The sample size ($n$) and the covariate dimension ($d$) are known factors at the design stage, whereas the correlation $\rho$ is an estimable factor that is also essentially known to the investigator.
Half of the $n$ units are assigned to the treatment group and the other half to the control group, where the allocation used is generated by one of the three randomization schemes
being compared. Given the covariates and allocation variable,
the outcome is simulated from the following model with a fixed additive treatment effect $\tau$,
\begin{align}\label{def:datgenlm}
y = g(\bx,\bbeta)+\tau W+\varepsilon,
\end{align}
where $g(\bx,\bbeta)$ represents the function of response surface that is unknown to the investigator but estimable to some extent, and the residual follows a normal distribution $\varepsilon\sim N(0,\sigma^2_\varepsilon)$, also unknown to the investigator. We adopt two different functions for the response surface: $g(\bx,\bbeta)\in\{\bx^\top\bbeta, \exp(\bx)^\top\bbeta\}$ with $\bbeta\in\big\{\bm{1}_d,(\bm{1}^\top_{d/2},2\times\bm{1}^\top_{d/2})^\top\big\}$.
The residual variance $\sigma_\varepsilon^2$ takes values from $\{0.5,1\}$, and we set $\tau=1$. All factors
with their levels and descriptions are summarized in Table \ref{table:factors}.

We compare PCA rerandomization (PCAReR) with three other randomization schemes: (a) complete (pure) randomization (CR) with the constraint (\ref{def: nt=nc}), (b) original rerandomization (ReR), and (c) ridge rerandomization (RidgeReR). 
For fair comparisons, we set the same acceptance probability for all rerandomization-based methods, i.e., $p_{a_k} = p_{a_\lambda} = p_a = 0.05$, corresponding to PCAReR, RidgeReR and ReR. For PCAReR, we set $\gamma_k = 0.95$ in (\ref{def:pcak}) to specify the number of top principal components to be selected. The optimal ridge coefficient $\lambda$ is determined by the procedure given in \cite{branson211ridge}.

All four randomization methods are evaluated using three criteria: the final covariate balance, the true estimation precision of an additive treatment effect, and computational time. Because all approaches produce balanced covariates on average, we adopt the
average variance of the mean difference $\bar{x}_{T,j}-\bar{x}_{C,j}$ across all covariates $j=1,\dots,d$, denoted by $\bar{{\sigma}}^2$, to evaluate the empirical balance. For the treatment effect, we choose the mean squared error (MSE) of $\hat{\tau}$ as the
evaluation metric. To normalize these metrics, we compute the reduction percents of $\bar{{\sigma}}^2$ and MSE for each rerandomization scheme relative to CR, which are denoted by $r_{\bar{{\sigma}}^2}$ and $r_{\rm MSE}$ respectively. We also record the computational time for generating a feasible allocation in seconds under each rerandomization method. It is important to have a fast algorithm if further analysis requires generating many acceptable allocations \citep{morgan2012rerandomization}, such as performing randomization tests.
In addition, we consider
the selected number of top components $k$ as well as the variance shrinkage coefficient $v_{a_k}$ in Theorem \ref{thm:cov_reduce} for PCAReR to help explain its performance under different settings.

Given a triplet of $(n, d,\rho)$, we simulate 2000 covariate matrices. For each covariate matrix,
four randomization schemes are used to generate allocations and then the responses
are simulated from model (\ref{def:datgenlm}) for all combinations of $\big\{g(\bx,\bbeta), \bbeta, \sigma^2_\varepsilon\big\}$.
Following \cite{rubin1979using},
we use the same covariate matrix when comparing different rerandomization methods, and the covariate matrices with smaller $n$ and $d$, such as $(n,d)=(100, 10)$, are correspondingly obtained from the first $100$ rows and $10$ columns of the covariate matrices with $(n,d)=(1000,180)$ given the same $\rho$. This nested design strategy minimizes the number of random samples and correlates the results of
different rerandomization methods, which thus makes comparisons more precise.

All evaluation metrics are calculated under each configuration of factors based on 2000 replications. We divide 2000 replications into 10 separate groups of size 200 to create repeated metrics for calculating their within-configuration mean square in ANOVA. In each group, we compute the three major metrics $r_{\bar{{\sigma}}^2}$, $r_{\rm MSE}$ and the computational time.
Under each rerandomization scheme,
we can obtain $4\times 4 \times 3 \times 10=480$ different values for both
$r_{\bar{\sigma}^2}$ and computational time based on our replication strategy, because these two metrics are completely determined by three known or estimable factors.  However, given the scheme,
we have $4\times 4 \times 3 \times 2^3 \times 10 = 3840$ different values
for $r_{\rm MSE}$ with respect to all the remaining six factors.

\subsection{Comparisons of Rerandomization Schemes}
Following \cite{rubin1979using} and \cite{gutman2013robust},
we begin with three separate ANOVAs
based on the seven-factor design to identify the most influential factors
in terms of three major evaluation metrics.
In the ANOVA, we consider  the main effects of the factors as well as all of their interactions.
The relative importance of factors and their interactions are evaluated by their $F$-ratios or, equivalently, the ANOVA mean squares.

Table \ref{table:anova_rsig} presents the ANOVA results of $r_{\bar{{\sigma}}^2}$ for known factors,
where we find that the covariate dimension, $d$, and correlation, $\rho$,
strongly influence the covariate balance metric, and the first three factors ($d$, scheme and $\rho$) account for around $82\%$ of the total sum of squares. We only report the top 11 factorial effects because the rest have mean squares much smaller than the mean squares for the top 11. Furthermore, the within-configuration mean square indicates small variability of the metric due to random sampling in the data generation.
The ANOVA results of $r_{\rm MSE}$ are reported in Table \ref{table:anova_rmse},
where we only display the top 20 influential factorial factors based on the $F$-ratios. We observe that the covariate dimension, $d$, the
correlation coefficient, $\rho$, and the type of response surface, $g(\bx,\bbeta)$, are
the most important factors in addition to the scheme, and the residuals also have relatively small mean square. Additionally, the first four factors explain around $72\%$ of the total sum of squares. From Table \ref{table:anova_time},
we can identify, the scheme, $d$ and $n$
as the most influential factors for the computational time, and about $68\%$ of the total sum of squares can be explained by the first five factorial effects.
After identifying the crucial factors, we take the average of each metric over the levels of all the remaining factors
to provide a comprehensive characterization for the effects of the selected factors.

\begin{figure}[tb]
\centering
\includegraphics[width=0.6\linewidth]{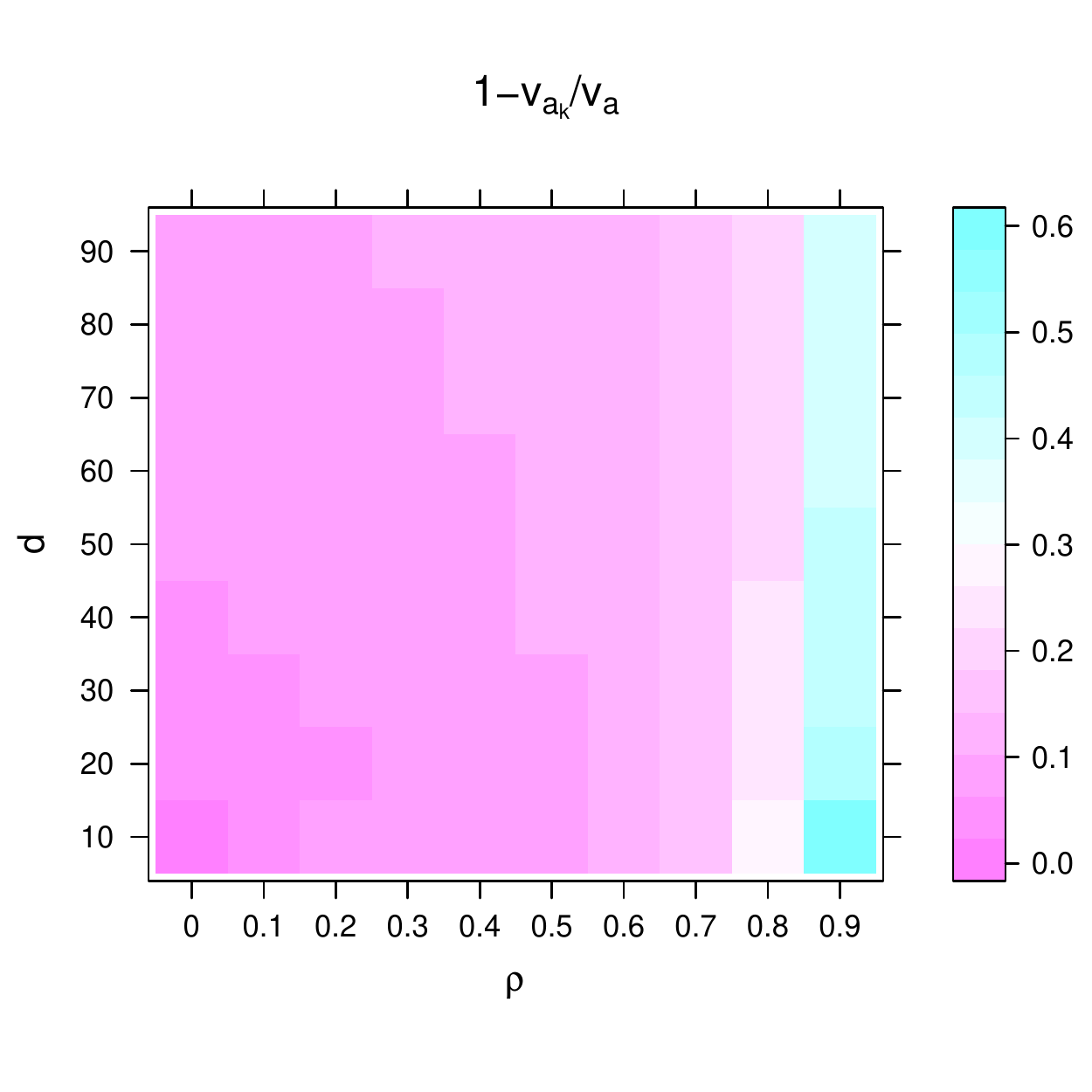}
\caption{The reduction percent of the variance shrinkage coefficient $v_{a_k}$
of PCAReR with respect to $v_a$ of ReR in terms of the feature dimension $d\in\{10,\dots,90\}$ and the correlation coefficient $\rho\in\{0,\dots,0.9\}$ with $n=100$.}
\label{fig:va_ratio}
\end{figure}

The first panel of Table \ref{table:sum_rsig_rmse} compares different rerandomization schemes in terms of $r_{\bar{{\sigma}}^2}$ with respect to $\rho$ and $d$.
To indicate the degree of dimension reduction by PCA,
we also note the number of selected top components
in parentheses.
In general, although PCAReR uses a smaller number of principal components, it can
result in more balanced covariates than ReR and the improvement is more obvious
in the cases with large $\rho$ and $d$. The main reason is that the variance shrinkage coefficient $v_{a_k}$ of PCAReR is typically smaller than the counterpart $v_a$ of ReR especially for large $\rho$ and $d$, as revealed in Figure \ref{fig:va_ratio}.
Compared with RidgeReR, which takes all principal components into account and puts more emphasis on the top ones,
PCAReR only uses the most variable subspace to perform the rerandomization,
and thus its performance is inferior to a certain extent.
The difference in $r_{\bar{{\sigma}}^2}$ between RidgeReR and PCAReR is small when $k$ is close to $d$,
for example, when $k=$ 9.9 or 43.2. 

Table \ref{table:sum_rsig_rmse} also presents $r_{\rm MSE}$ for each rerandomization scheme under different settings of the influential factors $(d,\rho,g(\bx,\bbeta))$. Because the estimation precision of $\hat{\tau}$ is closely related to the covariate balance as
implied by Theorem \ref{thm: tau_reduce}, we again observe that PCAReR has an intermediate performance between ReR and RidgeReR for a given type of response surface, and the advantages of PCAReR and RidgeReR are more evident for large $\rho$ and $d$. All methods yield smaller $r_{\rm MSE}$ for the nonlinear surface $\exp(\bx)^\top\bbeta$ relative to the linear surface $\bx^\top\bbeta$, because these approaches only balance the first moments of covariates. Additionally, all
three rerandomization schemes perform better for small $d$ given the same $\rho$ and $g(\bx,\bbeta)$, because it is more difficult to simultaneously balance numerous covariates.

As shown by Theorem \ref{thm: tau_reduce} in this paper, and Theorem 4.3 in \cite{branson211ridge}, the estimation precision of $\hat{\tau}$ for PCAReR and RidgeReR involves the transformed coefficient $\tilde{\bbeta}=\bV^\top\bbeta$. Therefore, it is expected a superior performance of PCAReR relative to RidgeReR, when the values of $\tilde{\bbeta}$ corresponding to the top components are smaller than those of the tail components among the selected $k$ principal components by PCA, such as $\tilde{\bbeta}=(1,1+2,\dots,\sum_{i=1}^ki,\bm{0}^\top_{d-k})^{\top}$. We demonstrate this phenomenon through a simulation by setting the coefficient as $\bbeta=\bV\tilde{\bbeta}$ under a linear response surface $g(\bx,\bbeta)=\bx^\top\bbeta$ with sample size $n=200$ and the residual variance $\sigma^2_\varepsilon=1$. The factors $d$ and $\rho$ are kept at the same values as given in Table \ref{table:factors}. The last panel of Table 
\ref{table:sum_rsig_rmse}
presents the percent reduction $r_{\rm MSE}$. We still observe that PCAReR works better than ReR, because PCAReR tends to yield more balanced covariates in the selected subspace. This example shows that PCAReR may outperform RidgeReR if the subspace spanned by the tail components among the $k$ selected ones is associated with larger values of regression coefficients.

Table \ref{table:sum_time} summarizes the computational time for three rerandomization schemes, which shows that PCAReR dominates in most cases. Generally, it can be complicated for RidgeReR to specify the optimal values for $\lambda$ and $a_\lambda$, so RidgeReR spends considerable time in generating a feasible allocation. When $d$ is close to $n$, such as $(n,d)=(100,90)$ or $(200,180)$,
it takes ReR much longer than other cases with $n\gg d$, to generate an allocation.
This is due to the long rejection period for a given threshold $a$, because it is more difficult to balance a large number of covariates.
Note that $a$ is determined under a multivariate normal assumption that only holds asymptotically, and the assumption may not be satisfied when $d$ is close to $n$, so the given threshold
level $a$ may be too restrictive for ReR. This result also reflects the practical
advantage of using PCA to accelerate the search for a feasible allocation.

In Table \ref{table:sum_time}, ReR displays some
unusual results when $d >n $, i.e., $(n,d)=(100,180)$.
In such a case, Mahalanobis distance of ReR reduces to a constant,
\begin{align*}
M = \frac{n-1}{n}(2\bW-\bm{1}_n)^\top\bX\left(\bX^\top\bX\right)^{-}\bX^\top(2\bW-\bm{1}_n)=n-1,
\end{align*}
where $\left(\bX^\top\bX\right)^{-}$ refers to the pseudo-inverse of
$\bX^\top\bX$ and $\bU\in\mathbb{R}^{n\times n}$ is composed of the left singular vectors
of $\bX$ with $\bX\left(\bX^\top\bX\right)^{-}\bX^\top=\bU\bU^\top=\bm{I}_n$. We further find that the threshold $a$ of ReR is larger than $n-1$ for $(n,d)=(100,180)$,
so that ReR is reduced to CR in such situations, which explains the computational times reported in Table \ref{table:sum_time}.


\subsection{Guideline for Rerandomization}
Based on the simulation results, we provide a guideline for
the investigator to select a proper scheme under different settings of factors. Tables \ref{table:anova_rsig}--\ref{table:anova_time} show that $d$ and $\rho$ are the most influential factors in terms of the $F$-ratios. Given the known factors $d$ and $n$ and estimable factor $\rho$, the investigator may choose a suitable rerandomization scheme as follows.
When both $d$ and $\rho$ are small,
all of the three rerandomization schemes show similar performance in terms of  $r_{\bar{{\sigma}}^2}$ and $r_{\rm MSE}$. When either $d$ or $\rho$ is large, PCAReR and RidgeReR should be considered because they leverage the importance of principal components to improve covariate balance and thus estimation precision. Note that $\bbeta$ and $g(\bx,\bbeta)$ are unknown to the investigator during the design stage, and 
Table \ref{table:sum_rsig_rmse}
implies that both methods could yield better performance than ReR. We hence recommend using PCAReR because
the fast implementation and sufficient dimension reduction of PCAReR
allows the investigator to make prompt attempts on
expanding the dimension of the covariate matrix by incorporating different combinations of nonlinear
features (e.g., second moments of covariates),
and may boost its performance
using nonlinear response surfaces.

\subsection{Real Application}\label{sec:realstudy}
For illustration, we compare different rerandomization methods using a real dataset. We use the first part of the data obtained from the Infant
Health and Development Program (IHDP) \citep{hill2011bayesian, louizos2017causal}.  This program aimed at improving cognitive developtment for the low-birth-weight and premature infants by providing high-quality child care and home visits from trained specialists for the infants in the treatment group. This intervention successfully promoted cognitive test scores for the treated children compared with the control group.  The dataset consists of 747 participants with 25 covariates and the continuous test scores. Six covariates are continuous which are standardized, and the others are binary.  To mimic the real process, we use all observed data to fit a sparse linear model via LASSO using 25 main effect covariates and all $ {25 \choose 2}=300$ two-way interactions among variables. The LASSO penalty is chosen to be 0.055 from cross-validation, and 35 covariates are finally selected. The fitted model is used to generate responses given the allocation during randomization.  We construct the covariate matrix $\bX\in\mathbb{R}^{747\times 325}$ by all instances, and plan to assign an approximately equal number of units to the treatment (373) and control (374) groups. Different randomization approaches are then applied to generate 1000 independent allocations, from which we calculate various empirical metrics.

\begin{figure}[tb]
\centering
\subfigure[Covariate Balance\label{fig:ihdp_meandiff}]{\includegraphics[width=.8\linewidth]{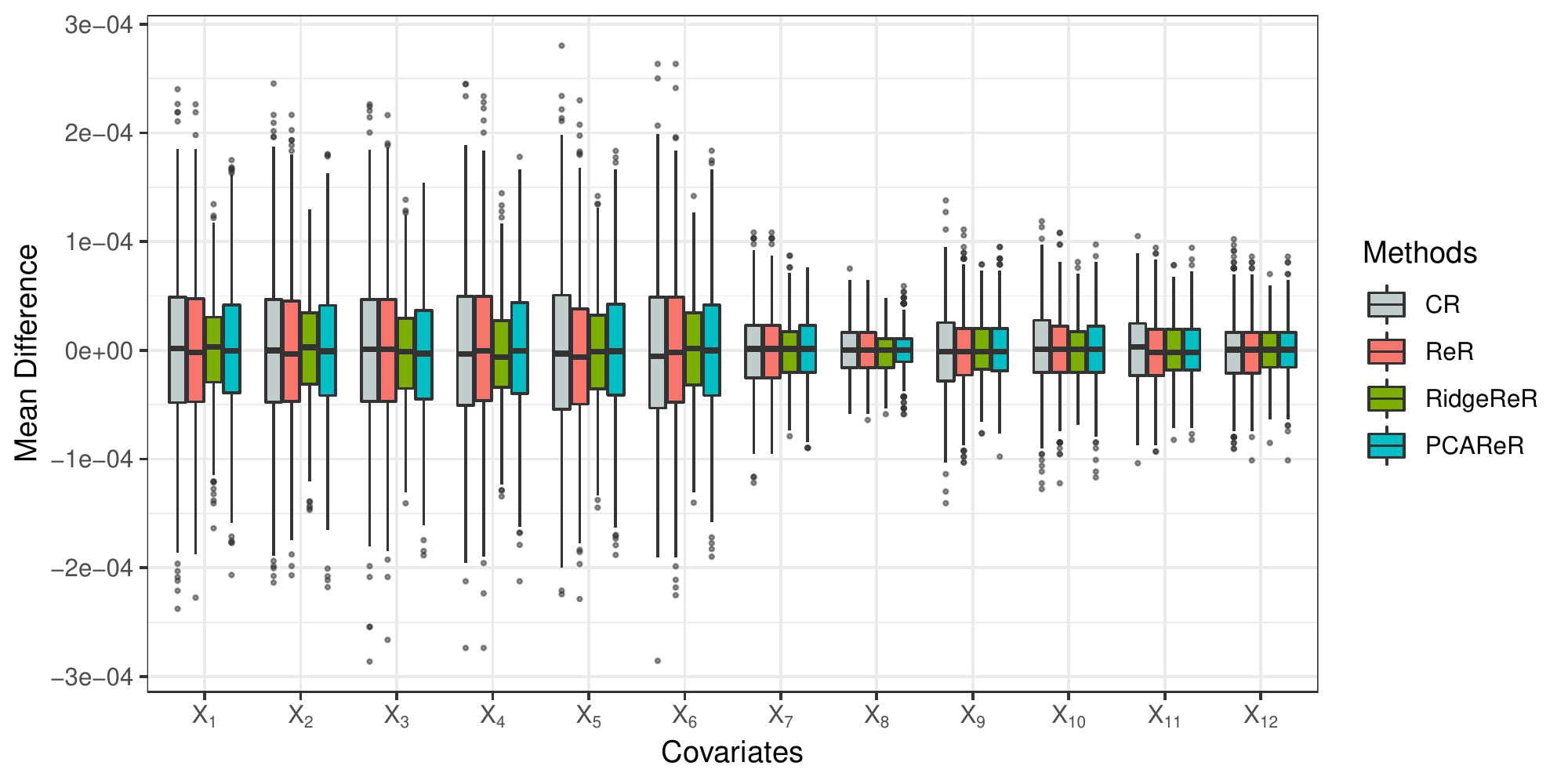}}
\subfigure[Density of $\hat{\tau}$\label{fig:ihdp_denplot}]{\includegraphics[width=.6\linewidth]{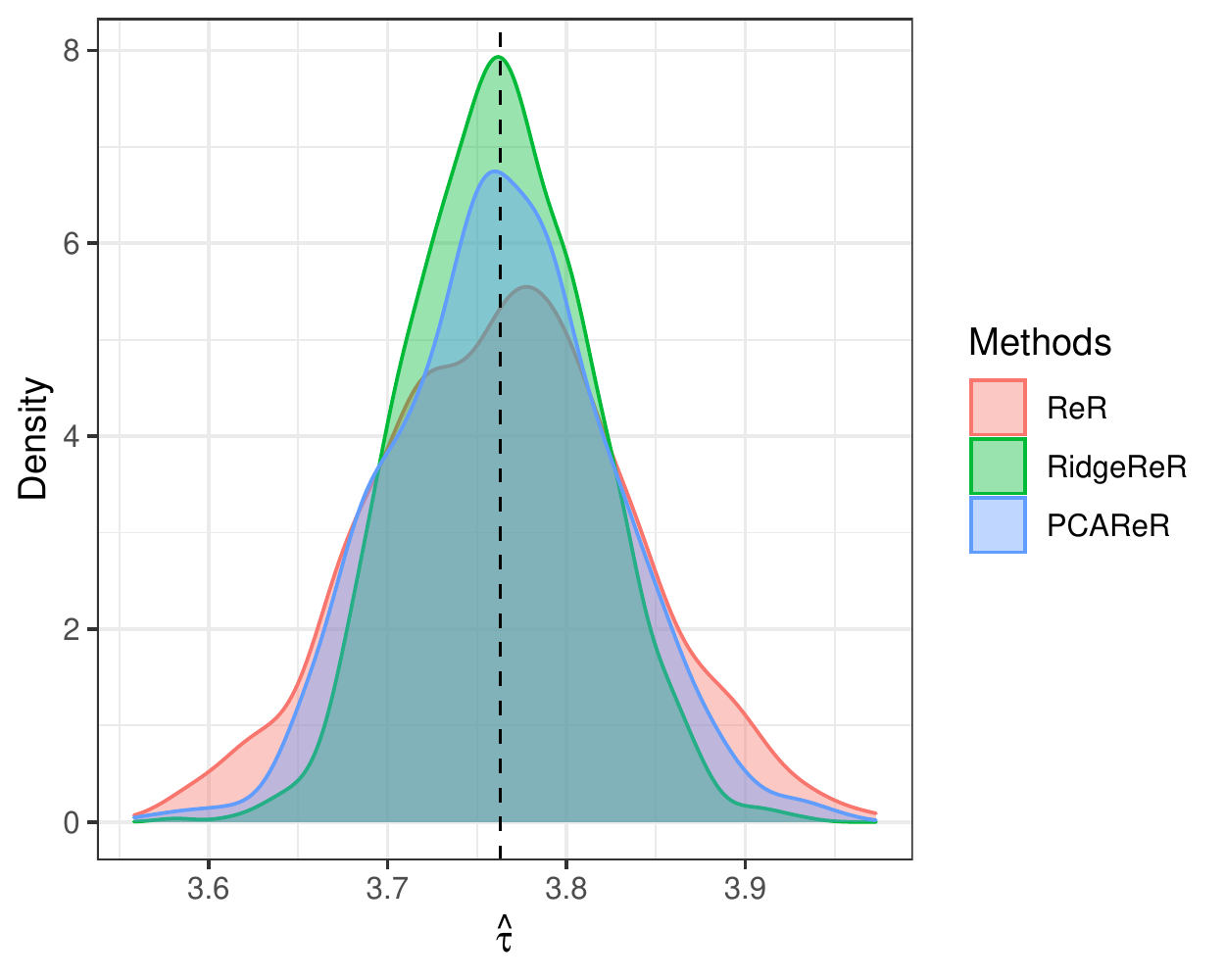}}
\caption{The results of different randomization approaches on the IHDP dataset. (a) Box plot of covariate mean differences between the treatment and control for continuous variables ($X_1,\dots,X_6$) and first six binary variables ($X_7,\dots,X_{12}$);
(b) density plots of the estimated treatment effect, where the black vertical line denotes the `true' value in the linear model. CR: Complete randomization with equal size of units in each group, ReR: Rerandomization, RidgeReR: ridge rerandomization, PCAReR: PCA rerandomization.}
\end{figure}

We first draw the box plot of the mean difference $\bar{x}_{T,j}-\bar{x}_{C,j}$ for continuous features ($X_1,\dots,X_6$) and first six binary variables ($X_7,\dots,X_{12}$) in Figure \ref{fig:ihdp_meandiff}; the other binary covariates demonstrate similar balance properties as the first six ones. We observe that PCAReR  using $81$ principal components achieves more balanced covariates than CR and its performance is comparable to ReR. We further present the density plot of the treatment effect estimator $\hat{\tau}$ with respect to 1000 replications for ReR, RidgeReR and PCAReR in Figure \ref{fig:ihdp_denplot}, where the distribution of $\hat{\tau}$ under PCAReR is more concentrated than that under ReR. Table \ref{table:ihdp} reports the overall reduction percentages for covariate mean differences and the MSE of $\hat{\tau}$ with respect to CR, where PCAReR yields performance between ReR and RidgeReR, but PCAReR runs 159 times faster than RidgeReR, which takes around 7.9 hours for a total of 1000 allocations, which could be an issue if inferences were to be based on randomization tests.

\section{Discussion}\label{sec:discuss}
We propose a PCA rerandomization scheme to allocate participants in experiments where there are high-dimensional or high-collinearity covariates. Compared with the classical rerandomization \citep{morgan2012rerandomization}, the major difference is that we use the top principal components rather than the original covariates to calculate Mahalanobis distance. We show that not only does PCA rerandomization share most theoretical characteristics with the simple rerandomization, but it also demonstrates empirical advantages with high-dimensional or highly correlated covariates in terms of covariate balance, the precision of treatment effect estimation as well as computational time.

Analogous to rerandomization, we can extend PCA rerandomization to other types of experiments. A line of future work is to develop the asymptotic property of the treatment effect estimator under PCA rerandomization as done in \cite{li2018asymptotic} for rerandomization. Another path worth exploring is to apply PCA rerandomization to $2^K$ factorial designs following \cite{branson2016improving}.

\clearpage
\bibliographystyle{apa}
\bibliography{pcarr}%

\clearpage
\appendix
\section{Appendix}

\subsection{Proof of Theorem \ref{thm:dist_Mpca}}
\begin{proof}
From (\ref{def: xt&xc}), we have
\begin{align*}
\bar{\bx}_T-\bar{\bx}_C = \frac{2}{n}\bX^{\top}(2\bW-\mathbf{1}_n)=\frac{2}{n}\bV\bZ^{\top}(2\bW-\mathbf{1}_n)\sim N(\mathbf{0},\bSigma).
\end{align*}
Furthermore, it is easy to see that
\begin{align}\label{eqn:ud&sigmat}
\bZ = (\bZ_k,\tilde{\bZ}_{d-k})\quad \mbox{and} \quad \bV^{\top}\bSigma\bV = C_n\bD^2=C_n
\begin{pmatrix}
\bD_k^2 & \mathbf{0}\\
\mathbf{0} & \tilde{\bD}_{d-k}^2
\end{pmatrix},
\end{align}
where $C_n=4/(n^2-n)$.
Consequently,
\begin{align}\label{eqn:diff_zt&zc}
\bar{\bz}^{(k)}_T-\bar{\bz}^{(k)}_C = \frac{2}{n}\bZ_k^{\top}(2\bW-\mathbf{1}_n)=\frac{2}{n}(\bI_k,\mathbf{0})\bV^{\top}\bX^{\top}(2\bW-\mathbf{1}_n) \sim N(\mathbf{0},\bSigma_{z}),
\end{align}
where $\bSigma_{z}=\bZ^{\top}_k\bZ_k=C_n\bD_{k}^2$. This concludes that $M_{k}|\bX\sim\chi^2_k$ according to the property of a multivariate normal distribution.
\end{proof}

\subsection{Proof of Theorem \ref{thm:cov_reduce}}
\begin{proof}
For simplicity, let $\bm{\eta}=\bV^{\top}(\bar{\bx}_T-\bar{\bx}_C)=(\bm{\eta}_k,\tilde{\bm{\eta}}_{d-k})^{\top}$ where $\bm{\eta}_k=(\eta_1,\dots,\eta_k)^\top\in\mathbb{R}^k$ and $\tilde{\bm{\eta}}_{d-k}=(\eta_{k+1},\dots,\eta_d)^\top\in\mathbb{R}^{d-k}$. According to (\ref{eqn:ud&sigmat}) and (\ref{eqn:diff_zt&zc}), we know that $\bm{\eta}|\bX\sim N(\mathbf{0},C_n\bD^2)$, $\bm{\eta}_k|\bX\sim N(\mathbf{0},C_n\bD^2_k)$ and $\tilde{\bm{\eta}}_{d-k}|\bX\sim N(\mathbf{0},C_n\tilde{\bD}^2_{d-k})$.
Thus, $M_{k}$ can be written in terms of $\bm{\eta}_k$ and
singular values $\sigma_1\geq \dots \geq \sigma_k>0$, i.e.,
$M_{k}=\sum_{j=1}^k\eta^2_j/(C_n\sigma_j^2)$.

Note that we have
\begin{align}\label{eqn:covxtxc_decomp}
\cov(\bar{\bx}_T-\bar{\bx}_C|\bX, M_{k}\leq a_k)=\bV\cov(\bm{\eta}|\bX, M_{k}\leq a_k)\bV^{\top}.
\end{align}
Let $a \overset{d}{=} b$ denote two variables $a$ and $b$ having the same distribution. First, one can obtain that  $\mathbb{E}(\eta_j|\bX,M_{k}\leq a_k)=0$ through the symmetry of 
a normal distribution, i.e.,  $-\eta_j\overset{d}{=}\eta_j\sim N(0, C_n\sigma^2_j), \forall j\in[k]$ \citep{branson211ridge}. Specifically,
\begin{align}\label{Eeta}
\mathbb{E}(\eta_j|\bX,M_{k}\leq a_k)
& =\mathbb{E}\left(\eta_j\bigg|\bX, \sum_{j=1}^k\frac{\eta^2_j}{C_n\sigma_j^2}\leq a_k\right)\nonumber\\
& = \mathbb{E}\left(-\eta_j\bigg|\bX, \sum_{j=1}^k\frac{(-\eta_j)^2}{C_n\sigma_j^2}\leq a_k\right)\nonumber\\
& = -\mathbb{E}(\eta_j|\bX,M_{k}\leq a_k),
\end{align}
which leads to $\mathbb{E}(\eta_j|\bX,M_{k}\leq a_k)=0$.
Now, we focus on the covariance, $\cov(\eta_i\eta_j|\bX,M_{k}\leq a_k)$ with $i\neq j$ and $i,j\in[k]$.
Similar to (\ref{Eeta}),
we only need to flip the sign of $\eta_j$ to show $\mathbb{E}(\eta_i\eta_j|\bX,M_{k}\leq a_k)=0$ and
thus further derive 
$\cov(\eta_i\eta_j|\bX,M_{k}\leq a_k)=\mathbb{E}(\eta_i\eta_j|\bX,M_{k}\leq a_k)=0$.
On the other hand, $\eta_j^2/(C_n\sigma_j^2)$ are independent and identically follows $\chi^2_1$ for $j\in[k]$. Combining the above results, the variance of $\eta_j$ for $j\in[k]$ is
\begin{align*}
{\rm var}(\eta_j|\bX, M_{k}\leq a_k) &= \mathbb{E}(\eta_j^2|\bX, M_{k}\leq a_k)\\
& = C_n\sigma_j^2 \mathbb{E}\left(\frac{\eta_j^2}{C_n\sigma_j^2}\bigg|\bX, \sum_{j=1}^k\frac{\eta^2_j}{C_n\sigma_j^2}\leq a_k\right)\\
& =\frac{C_n\sigma_j^2}{k}\mathbb{E}(M_{k}|\bX, M_{k}\leq a_k),
\end{align*}
where the last equation follows the exchangeability of i.i.d. $\{\eta_j^2/(C_n\sigma_j^2)\}_{j\in[k]}$. 
Because $M_{k}|\bX \sim \chi^2_k$, we have $\mathbb{E}(M_{k}|\bX, M_{k}\leq a_k)/k=v_{a_k}=P(\chi^2_{k+2}\leq a_k)/P(\chi^2_{k}\leq a_k)$ following the proof of Theorem 3.1 in \cite{morgan2012rerandomization}. Therefore, the variance is ${\rm var}(\eta_j|\bX, M_{k}\leq a_k)=C_nv_{a_k}\sigma_j^2$.

It is evident that for $j=k+1,\dots,d$, we have
\begin{align*}
\mathbb{E}(\eta_j|\bX,M_{k}\leq a_k)
&=\mathbb{E}(\eta_j|\bX)=0,\\
{\rm var}(\eta_j|\bX,M_{k}\leq a_k)
&={\rm var}(\eta_j|\bX)=C_n\sigma_j^2,\\
\cov(\eta_i\eta_j|\bX,M_{k}\leq a_k)
&= \cov(\eta_i\eta_j|\bX) = 0, ~\forall i\neq j, ~ i=k+1,\dots,d.
\end{align*}
Finally, we show $\cov(\eta_i\eta_j|\bX,M_{k}\leq a_k)=0$ when $i\in[k]$ and $j=k+1,\dots,d$ as follows,
\begin{align*}
\cov(\eta_i\eta_j|\bX,M_{k}\leq a_k)
&= \mathbb{E}(\eta_i\eta_j|\bX,M_{k}\leq a_k)\\
&= \mathbb{E}\big\{\eta_i\mathbb{E}(\eta_j|\eta_i,\bX,M_{k}\leq a_k)|\bX,M_{k}\leq a_k\big\} \\
& = \mathbb{E}\big\{\eta_i\mathbb{E}(\eta_j|\bX)|\bX,M_{k}\leq a_k\big\}\\
&= 0.
\end{align*}
Considering all the aforementioned results of the variance and covariance, we have
$$\cov(\bm{\eta}|\bX,M_{k}\leq a_k)=C_n\begin{pmatrix}
v_{a_k}\bD^2_k & \mathbf{0}\\
\mathbf{0} & \tilde{\bD}^2_{d-k}
\end{pmatrix}.$$
The theorem can be proved after plugging the above equation into (\ref{eqn:covxtxc_decomp}).
\end{proof}

\subsection{Proof of Corollary \ref{cor: cov_reduce}}
\begin{proof}
According to the definition of PRV, we can obtain the expression of $v_{a_{k,j}}$ as the ratio of the $j$th diagonal entry of the covariance matrix for PCAReR and $\bSigma$; that is,
$$v_{a_k,j} = \frac{\big(C_n\bV{\rm diag}\big\{v_{a_k}\bD^2_k,\tilde{\bD}^2_{d-k}\big\}\bV^{\top}\big)_{jj}}{\bSigma_{jj}}.$$
Here, we show that $v_{a_k,j}\in(0,1)$.
Define $\bV=(\bV_k,\tilde{\bV}_{d-k})$ where $\bV_k\in\mathbb{R}^{d\times k}$ and $\tilde{\bV}_{d-k}\in\mathbb{R}^{d\times (d-k)}$ correspond to 
the first $k$ and last $d-k$ components. Let $\balpha_j=(\mathbf{0}^\top_{j-1},1,\mathbf{0}^\top_{d-j})^\top$ be a $d$-dimensional
vector of 0's except for the $j$th element taking a value of 1.
We have that
\begin{align*}
 \bSigma_{jj}
& = \balpha_j^{\top}\bSigma\balpha_j \\
&= C_n\balpha_j^{\top}\bV
\begin{pmatrix}
\bD^2_k & \mathbf{0}\\
\mathbf{0} & \tilde{\bD}^2_{d-k}
\end{pmatrix}\bV^{\top}\balpha_j\\
&=C_n\balpha_j^{\top}\bV_k\bD_k^2\bV_k^{\top}\balpha_j+C_n\balpha_j^{\top}\tilde{\bV}_{d-k}\tilde{\bD}_{d-k}^2\tilde{\bV}_{d-k}^{\top}\balpha_j>0,
\end{align*}
and
\begin{align*}
\big(C_n\bV\diag\big\{v_{a_k}\bD^2_k,\tilde{\bD}^2_{d-k}\big\}\bV^{\top}\big)_{jj}  = v_aC_n\balpha_j^{\top}\bV_k\bD_k^2\bV_k^{\top}\balpha_j+C_n\balpha_j^{\top}\tilde{\bV}_{d-k}\tilde{\bD}_{d-k}^2\tilde{\bV}_{d-k}^{\top}\balpha_j.
\end{align*}
Since $v_{a_k} = P(\chi^2_{k+2}\leq a_k)/P(\chi^2_{k}\leq a_k)\in(0,1)$ for any $a_k>0$. We can conclude that
$$0<\big(C_n\bV\diag\big\{v_{a_k}\bD^2_k,\tilde{\bD}^2_{d-k}\big\}\bV^{\top}\big)_{jj}< \bSigma_{jj},$$
and thus $v_{a_{k,j}}\in(0,1)$
because both $\bV_k\bD_k^2\bV_k^{\top}$ and $\tilde{\bV}_{d-k}\tilde{\bD}_{d-k}^2\tilde{\bV}_{d-k}^{\top}$ are positive definitive.
\end{proof}

\subsection{Proof of Theorem \ref{thm: tau_reduce}}
\begin{proof}
Define $\bar{\varepsilon}_{T}=2\bW^{\top}\bvarepsilon/n$ and
$\bar{\varepsilon}_{C}=2(\mathbf{1}-\bW)^{\top}\bvarepsilon/n$, where
$\bvarepsilon=(\varepsilon_1,\dots,\varepsilon_n)^{\top}$.
According to (\ref{def:taumodel}), $\hat{\tau}$ can be written as
$$\hat{\tau} = (\bar{\bx}_T-\bar{\bx}_C)^{\top}\bbeta+\tau+(\bar{\varepsilon}_T-\bar{\varepsilon}_C).$$
Leveraging the orthogonality between the first and last terms, we have that
\begin{align}\label{eqn:vartau_x}
{\rm var}(\hat{\tau}|\bX) &= {\rm var}\left\{(\bar{\bx}_T-\bar{\bx}_C)^{\top}\bbeta|\bX\right\} + {\rm var}(\bar{\varepsilon}_T-\bar{\varepsilon}_C|\bX)\nonumber\\
&= \bbeta^{\top}\bSigma\bbeta + {\rm var}(\bar{\varepsilon}_T-\bar{\varepsilon}_C|\bX)\nonumber\\
&=  C_n\bbeta^{\top}\bV\begin{pmatrix}
\bD^2_k & \mathbf{0}\\
\mathbf{0} & \tilde{\bD}^2_{d-k}
\end{pmatrix}\bV^{\top}\bbeta + {\rm var}(\bar{\varepsilon}_T-\bar{\varepsilon}_C|\bX).
\end{align}
Furthermore, the conditional normal assumption on $\hat{\tau}$ and $\bar{\bx}_T-\bar{\bx}_C$ leads to the conditional independence between $\bar{\bx}_T-\bar{\bx}_C$ and $\bar{\varepsilon}_T-\bar{\varepsilon}_C$, because these two terms are uncorrelated. Therefore, $M_{k}$ is also conditionally independent of $\bar{\varepsilon}_T-\bar{\varepsilon}_C$, and we have that
\begin{align}\label{eqn:vartau_Mpca}
{\rm var}(\hat{\tau}|\bX, M_{k} \leq a_k) &= {\rm var}\left\{(\bar{\bx}_T-\bar{\bx}_C)^{\top}\bbeta|\bX, M_{k} \leq a_k\right\} + {\rm var}(\bar{\varepsilon}_T-\bar{\varepsilon}_C|\bX, M_{k} \leq a_k)\nonumber\\
&= \bbeta^{\top}\cov(\bar{\bx}_T-\bar{\bx}_C|\bX, M_{k} \leq a_k)\bbeta + {\rm var}(\bar{\varepsilon}_T-\bar{\varepsilon}_C|\bX)\nonumber\\
&= C_n\bbeta^{\top}\bV\begin{pmatrix}
v_{a_k}\bD^2_k & \mathbf{0}\\
\mathbf{0} & \tilde{\bD}^2_{d-k}
\end{pmatrix}\bV^{\top}\bbeta + {\rm var}(\bar{\varepsilon}_T-\bar{\varepsilon}_C|\bX).
\end{align}
Combining (\ref{eqn:vartau_x}) and (\ref{eqn:vartau_Mpca}), it can be shown that
$${\rm var}(\hat{\tau}|\bX) - {\rm var}(\hat{\tau}|\bX, M_{k} \leq a_k) = C_n\bbeta^{\top}\bV\begin{pmatrix}
(1-v_{a_k})\bD^2_k & \mathbf{0}\\
\mathbf{0} & \mathbf{0}
\end{pmatrix}\bV^{\top}\bbeta\geq0,$$
where the non-negativity arises from the positive semi-definiteness of the matrix. When $\bbeta\in\mathcal{C}$, since $v_{a_k}\in(0,1)$ and $\bD_k$ is positive definite, one can conclude that ${\rm var}(\hat{\tau}|\bX) > {\rm var}(\hat{\tau}|\bX, M_{k} \leq a_k)$.
\end{proof}

\clearpage

\clearpage

\newpage

\begin{table}[]
\centering
\caption{Factors and their corresponding levels in the simulation study.}
\label{table:factors}
\begin{threeparttable}
\begin{tabular}{@{}lll@{}}
\toprule
Factors  & Levels & Descriptions   \\
\midrule
$n$     & $\{100, 200, 500, 1000\}$ & Sample size \\
$d$      & $\{10, 50, 90,180\}$ & Dimension of covariates \\
$\rho$ & $\{0.1, 0.5, 0.9\}$ & Correlation coefficient of covariates \\
Scheme & $\{\mbox{ReR}, \mbox{RidgeReR}, \mbox{PCAReR}\}$ & Rerandomization (ReR) scheme\\
$g(\bx,\bbeta)$   & $\{\bx^\top\bbeta, \exp(\bx)^\top\bbeta\}$  & Response surface  \\
$\bbeta$ & $\big\{\bm{1}_d,(\bm{1}^\top_{d/2},2\times \bm{1}^\top_{d/2})^\top\big\}$ & Coefficient vector in the response surface \\
$\sigma^2_{\varepsilon}$ & $\{0.5, 1\}$ & Residual variance \\
\bottomrule
\end{tabular}
\begin{tablenotes}[flushleft]
\item
\end{tablenotes}
\end{threeparttable}
\end{table}

\begin{table}[]
\centering
\caption{ANOVA results of the 11 most influential factors
for $r_{\bar{{\sigma}}^2}$,
which is the reduction percent of the average empirical variance of $\bar{x}_{T,j}-\bar{x}_{C,j}$ across
all covariates ($j=1,\dots,d$) relative to complete randomization. DF is the degree of freedom, MS represents mean square, and the last row Residual refers to the within-configuration mean square.}
\label{table:anova_rsig}
\begin{threeparttable}
\begin{tabular}{@{}lrrr@{}}
\toprule
Factors        & DF   & MS & $F$-ratio \\ \midrule
$d$              & 3    & 13.31       & 7444   \\
$\mbox{Scheme}$         & 2    & 5.72        & 3201   \\
$\rho$            & 2    & 3.91        & 2185  \\
$\rho\times \mbox{scheme}$     & 4    & 1.41        & 789    \\
$d\times \mbox{scheme}$       & 6    & 0.28        & 157    \\
$n\times \mbox{scheme}$       & 6    & 0.09        & 50     \\
$d\times \rho$          & 6    & 0.06        & 32     \\
$n\times \rho\times \mbox{scheme}$   & 12   & 0.04        & 25     \\
$d\times \rho\times \mbox{scheme}$   & 12   & 0.04        & 21     \\
$n\times d\times \mbox{scheme}$     & 18   & 0.03        & 19    \\
$n$              & 3    & 0.03        & 15     \\
\midrule
Residual      & 1296 & 0.002       &           \\ \bottomrule
\end{tabular}
\end{threeparttable}
\end{table}

\begin{table}[]
\centering
\caption{ANOVA results of the 20 most influential factors based on the ANOVA $F$-ratio
for $r_{\rm MSE}$, which is the reduction percent of mean squared error (MSE)
of $\hat{\tau}$ relative to complete randomization. DF is the degree of freedom, MS represents mean square, and the last row Residual refers to the within-configuration mean square.}
\label{table:anova_rmse}
\begin{threeparttable}
\begin{tabular}{@{}lrrr@{}}
\toprule
Factors          & DF    & MS & $F$-ratio \\ \midrule
$\mbox{Scheme}$           & 2     & 74.83       & 12128  \\
$d$                & 3     & 56.67       & 9184   \\
$g(\bx,\bbeta)$            & 1     & 55.71       & 9029   \\
$\rho$              & 2     & 5.85        & 949    \\
$\rho\times g(\bx,\bbeta)$        & 2     & 5.78        & 937    \\
$d\times \mbox{scheme}$         & 6     & 3.47        & 563    \\
$\rho\times \mbox{scheme}$       & 4     & 3.43        & 556    \\
$d\times g(\bx,\bbeta)$          & 3     & 3.06        & 496    \\
$\mbox{Scheme}\times g(\bx,\bbeta)$    & 2     & 1.96        & 317   \\
$n\times \mbox{scheme}$         & 6     & 1.24        & 201    \\
$n$                & 3     & 0.63        & 103    \\
$\rho\times \mbox{scheme}\times g(\bx,\bbeta)$ & 4     & 0.43        & 69     \\
$n\times d\times \mbox{scheme}$       & 18    & 0.27        & 44    \\
$d\times \rho$           & 6     & 0.22        & 35     \\
$n\times \rho\times \mbox{scheme}$     & 12    & 0.20        & 32     \\
$d\times \rho\times \mbox{scheme}$     & 12    & 0.09        & 15     \\
$n\times \rho$            & 6     & 0.09        & 15     \\
$d\times \mbox{scheme}\times g(\bx,\bbeta)$   & 6     & 0.08        & 14    \\
$n\times g(\bx,\bbeta)$          & 3     & 0.05        & 7      \\
$n\times d$              & 9     & 0.04        & 7      \\
\midrule
Residual        & 10368 & 0.006       &           \\ \bottomrule
\end{tabular}
\end{threeparttable}
\end{table}

\begin{table}[]
\centering
\caption{ANOVA results of known or estimable factors for the computational
time of each rerandomization scheme in seconds. DF is the degree of freedom, MS represents mean square, and the last row Residual refers to the within-configuration mean square.}
\label{table:anova_time}
\begin{threeparttable}
\begin{tabular}{@{}lrrr@{}}
\toprule
Factors        & DF   & MS & $F$-ratio \\ \midrule
$d$              & 3    & 42379     & 5363    \\
$\mbox{Scheme} $        & 2    & 18425     & 2332    \\
$n\times d\times \mbox{scheme} $    & 18   & 16566     & 2096    \\
$n\times \mbox{scheme}$       & 6    & 13658     & 1728    \\
$n$              & 3    & 11011    & 1393    \\
$d\times \mbox{scheme}$       & 6    & 10328     & 1307    \\
$n\times d$            & 9    & 9866      & 1249    \\
$\rho$            & 2    & 2959      & 374     \\
$d\times \rho $         & 6    & 2166      & 274     \\
$n\times \rho   $       & 6    & 1085      & 137    \\
$n\times d\times \rho\times \mbox{scheme} $& 36   & 1077     & 136     \\
$n\times \rho\times \mbox{scheme} $  & 12   & 1047      & 132     \\
$n\times d\times \rho$        & 18   & 851       & 108    \\
$\rho\times \mbox{scheme}$     & 4    & 748      & 95     \\
$d\times \rho\times \mbox{scheme}$   & 12   & 618      & 78      \\
\midrule
Residual      & 1296 & 8       &           \\ \bottomrule
\end{tabular}
\end{threeparttable}
\end{table}

\begin{landscape}
\begin{table}[]
\centering
\caption{The evaluation metrics $r_{\bar{{\sigma}}^2}\times 100$ and  $r_{\rm MSE}\times 100$ for three rerandomization (ReR)
schemes under different combinations of $\rho$ and $d$,
where $r_{\bar{{\sigma}}^2}$ and $r_{\rm MSE}$ denote the reduction percents of the average empirical variance of
$\bar{x}_{T,j}-\bar{x}_{C,j}$ across
all covariates ($j=1,\dots,d$) and mean squared error (MSE) of $\hat{\tau}$ relative to complete randomization, respectively. We report the number $k$ of selected top principal components
for PCA rerandomization in the parentheses. }
\label{table:sum_rsig_rmse}
\scalebox{1}{
\begin{tabular}{@{}lllllllllllll@{}}
\toprule
 & \multicolumn{3}{c}{$d=10$} & \multicolumn{3}{c}{$d=50$} & \multicolumn{3}{c}{$d=90$} & \multicolumn{3}{c}{$d=180$} \\   \cmidrule(l){2-4} \cmidrule(l){5-7} \cmidrule(l){8-10} \cmidrule(l){11-13}
Scheme   & $\rho=0.1$ & 0.5   & 0.9   & 0.1   & 0.5   & 0.9   & 0.1   & 0.5   & 0.9   & 0.1    & 0.5    & 0.9   \\
\midrule
         & \multicolumn{12}{c}{$r_{\bar{{\sigma}}^2}\times 100$}                                                                        \\
ReR      & 69        & 69  & 69 & 36    & 37    & 35   & 27    & 27    & 26   & 14     & 14    & 15   \\
RidgeReR & 70        & 76  & 90 & 40    & 51    & 79   & 33    & 45    & 74   & 28     & 40    & 72   \\
$\mbox{PCAReR}_{(k)}$  & $69_{(10)}$        & $69_{(9)}$  & $81_{(5)}$ & $37_{(43)}$    & $38_{(39)}$    & $54_{(17)}$   & $28_{(72)}$    & $30_{(65)}$    & $45_{(27)}$   & $21_{(123)}$     & $23_{(109)}$    & $38_{(43)}$   \\

         &             &       &      &         &         &        &         &         &        &          &         &        \\
         & \multicolumn{12}{c}{$r_{\rm MSE}\times 100$ for linear response surface}                                                     \\
ReR      & 65        & 68  & 69 & 36    & 37    & 35   & 27    & 28    & 26   & 14     & 13    & 15   \\
RidgeReR & 72        & 82  & 92 & 54    & 64    & 84   & 51    & 58    & 79   & 48     & 55    & 77   \\
PCAReR   & 66        & 71  & 85 & 38    & 40    & 57   & 30    & 32    & 48   & 25     & 25    & 41   \\
         &             &       &      &         &         &        &         &         &        &          &         &        \\
         & \multicolumn{12}{c}{$r_{\rm MSE}\times 100$ for nonlinear response surface}                                                  \\
ReR      & 47        & 50  & 41 & 32    & 29    & 20   & 25    & 22    & 16   & 14     & 10    & 11   \\
RidgeReR & 52        & 60  & 58 & 46    & 51    & 51   & 44    & 43    & 49   & 44     & 44    & 49   \\
PCAReR   & 47        & 54  & 53 & 33    & 32    & 38   & 26    & 26    & 26   & 23     & 20    & 23   \\
         &             &       &      &         &         &        &         &         &        &          &         &        \\
         & \multicolumn{12}{c}{$r_{\rm MSE}\times 100$ for a special choice of $\bbeta$}                                                \\
ReR      & 68        & 70  & 66 & 36    & 37    & 32   & 22    & 21    & 20   & 15     & 20    & 15   \\
RidgeReR & 68        & 65  & 70 & 28    & 36    & 39   & 21    & 20    & 38   & 15     & 17    & 35   \\
PCAReR   & 69        & 73  & 83 & 40    & 42    & 54   & 26    & 27    & 47   & 23     & 30    & 39   \\ \bottomrule
\end{tabular}
}
\end{table}
\end{landscape}

\begin{table}[]
\centering
\caption{The computational time in seconds for three rerandomization (ReR) schemes under different combinations of $n$ and $d$. The number of selected top principal components $k$
for PCA rerandomization is given in the parentheses.
}\label{table:sum_time}
\scalebox{1}{
\begin{tabular}{@{}lllllllll@{}}
\toprule
         & \multicolumn{4}{c}{$n=100$}   & \multicolumn{4}{c}{$n=200$}   \\
         \cmidrule(l){2-5} \cmidrule(l){6-9}
Scheme   & $d=10$ & 50   & 90    & 180   & 10   & 50   & 90   & 180    \\
\midrule
ReR      & 0.02 & 0.16 & 21.79 & 0.01  & 0.02 & 0.08 & 0.32 & 128.29 \\
RidgeReR & 1.01 & 3.24 & 5.94  & 92.65 & 1.41 & 2.99 & 6.06 & 19.41  \\
$\mbox{PCAReR}_{(k)}$   & $0.01_{(8)}$ & $0.05_{(29)}$ & $0.11_{(41)}$  & $0.62_{(55)}$  & $0.01_{(8)}$ & $0.04_{(33)}$ & $0.10_{(52)}$ & $0.43_{(82)}$   \\
\midrule
         & \multicolumn{4}{c}{$n=500$}   & \multicolumn{4}{c}{$n=1000$}  \\
          \cmidrule(l){2-5} \cmidrule(l){6-9}
Scheme         & $d=10$ & 50   & 90    & 180   & 10   & 50   & 90   & 180    \\
        \midrule
ReR      & 0.02 & 0.08 & 0.18  & 0.87  & 0.02 & 0.08 & 0.18 & 0.63   \\
RidgeReR & 1.45 & 2.97 & 5.73  & 16.19 & 1.21 & 3.15 & 6.32 & 19.79  \\
$\mbox{PCAReR}_{(k)}$   & $0.01_{(8)}$ & $0.05_{(36)}$ & $0.10_{(60)}$  & $0.32_{(109)}$  & $0.02_{(8)}$ & $0.05_{(37)}$ & $0.12_{(64)}$ & $0.36_{(120)}$   \\
\bottomrule
\end{tabular}
}
\end{table}

\begin{table}[]
\centering
\caption{Comparisons of different rerandomization (ReR) schemes on the real dataset, with $r_{\bar{{\sigma}}^2}$: the reduction percent of the variance of $\bar{x}_{T,j}-\bar{x}_{C,j}$ across
all covariates with respect to complete randomization (CR), $r_{\rm MSE}$: the reduction percent of mean squared error (MSE) of $\hat{\tau}$ relative to CR, and Time(s): computational time in seconds. }
\label{table:ihdp}
\begin{threeparttable}
\begin{tabular}{@{}lccc@{}}
\toprule
Scheme  & $r_{\bar{{\sigma}}^2}$& $r_{\rm MSE}$   & Time(s)       \\ \midrule
ReR      & 0.07 & 0.08 & 0.21 \\
RidgeReR & 0.26 & 0.56 & 28.54  \\
PCAReR ($k=81$)   & 0.21 & 0.34 & 0.18 \\ \bottomrule
\end{tabular}
\end{threeparttable}
\end{table}

\end{document}